\documentclass[conference]{IEEEtran}
\IEEEoverridecommandlockouts
\usepackage{amsmath,amssymb,amsfonts}
\usepackage{algorithmic}
\usepackage{graphicx}
\usepackage{textcomp}
\usepackage{xcolor}
\usepackage{hyperref}

\def\BibTeX{{\rm B\kern-.05em{\sc i\kern-.025em b}\kern-.08em
    T\kern-.1667em\lower.7ex\hbox{E}\kern-.125emX}}

\usepackage{multirow}
\usepackage{float}
\usepackage{hyperref}

\begin{document}

\title{A Data Science Approach for Detecting Honeypots in Ethereum}

\author{\IEEEauthorblockN{Ramiro Camino, Christof Ferreira Torres, Mathis Baden and Radu State}
\IEEEauthorblockA{\textit{SnT, University of Luxembourg, Luxembourg}\\
\{ramiro.camino,christof.torres,mathis.steichen,radu.state\}@uni.lu}
}

\maketitle

\begin{abstract}
Ethereum smart contracts have recently drawn a considerable amount of attention from the media, the financial industry and academia.
With the increase in popularity, malicious users found new opportunities to profit by deceiving newcomers.
Consequently, attackers started luring other attackers into contracts that seem to have exploitable flaws, but that actually contain a complex hidden trap that in the end benefits the contract creator.
In the blockchain community, these contracts are known as honeypots.
A recent study presented a tool called \textsc{HoneyBadger} that uses symbolic execution to detect honeypots by analyzing contract bytecode.
In this paper, we present a data science detection approach based foremost on the contract transaction behavior.
We create a partition of all the possible cases of fund movements between the contract creator, the contract, the transaction sender and other participants.
To this end, we add transaction aggregated features, such as the number of transactions and the corresponding mean value and other contract features, for example compilation information and source code length.
We find that all aforementioned categories of features contain useful information for the detection of honeypots.
Moreover, our approach allows us to detect new, previously undetected honeypots of already known techniques.
We furthermore employ our method to test the detection of unknown honeypot techniques by sequentially removing one technique from the training set.
We show that our method is capable of discovering the removed honeypot techniques.
Finally, we discovered two new techniques that were previously not known.

\end{abstract}

\section{Introduction}


Blockchain technology revolutionized the way assets are traded around the world.
A blockchain can be seen as an append-only ledger that is maintained by a decentralised peer-to-peer network.
The transfer of assets is recorded in form of transactions, which are further grouped into so-called \emph{blocks}.
Each block is linked to its predecessor via cryptographic means, thereby forming a ``blockchain''.
This list of blocks is stored across the participants of the blockchain.
A consensus protocol dictates the appending of new blocks.
Bitcoin~\cite{nakamoto2008bitcoin} and Ethereum~\cite{wood2014ethereum} are currently the two most prominent blockchain implementations.
Blockchain is considered as one of today's major innovations due to the fact that it allows to decentralize the governance of any asset.
While the purpose of the Bitcoin blockchain is to decentralize the governance of cash, and thereby the role of banks, the purpose of the Ethereum blockchain is to decentralize the computer as a whole.

Ethereum builds on top of Bitcoin's principles and additionally offers the execution of so-called \emph{smart contracts}.
Smart contracts are Turing-complete programs that are executed by the participants of the Ethereum peer-to-peer network.
They are identified by an address,
and are deployed or invoked via transactions. 
We distinguish between normal and internal transactions.
Normal transactions refer to transactions that are initiated by user accounts, whereas internal transactions refer to transactions that are initiated by smart contracts and are performed as a reaction to a normal transaction.
In other words, every internal transaction has a normal transaction as its origin.
Smart contracts may interact with Ethereum's native cryptocurrency called \emph{ether}, and receive data input from participants.
The blockchain enforces that smart contracts are run exactly as they are programmed, without any possibility of censorship. 
The execution of smart contracts costs \emph{gas}.
Users can determine how much gas they want to provide to the execution of a smart contract (gas limit), and how much they want to pay per gas unit in terms of ether (gas price).
If the total amount of gas assigned to a transaction is consumed before the end of the execution, the latter is terminated by throwing an error and reversing the effects of the transaction.
Smart contracts are usually programmed using a high-level programming language, such as Solidity~\cite{solidity}.
As with any program, smart contracts may contain bugs~\cite{atzei2017}.
Several smart contracts have already been attacked in the past.
The two most prominent hacks, the DAO hack~\cite{siegel2016understanding} and the Parity wallet hacks~\cite{petrov2017another}, caused together a loss of over \$400 million.
As a result, attackers started looking for vulnerable smart contracts \cite{atzei2017}.
Interestingly, recent works show that attackers started purposefully deploying vulnerable-looking smart contracts in order to lure other attackers into traps~\cite{ferreira2019art,etherwars}. 
This new type of fraud is known as ``honeypots''~\cite{hacking_the_hackers,honeypot_analysis}.
Honeypots are smart contracts that appear to have a flaw in their design.
This flaw typically allows any arbitrary user on the Ethereum blockchain to drain the funds contained in the contract.
However, once a user tries to exploit this apparent vulnerability, a trapdoor obfuscated into the contract prevents the draining of funds to succeed.
The attempt at exploitation of these honeypot contracts does not come for free.
Their exploitation is always linked to the transfer of a certain amount of ether to the contract.
The idea is that the victim solely focuses on the apparent vulnerability and does not consider the possibility that a trap might be hidden inside the contract.
Moreover, honeypot creators often upload the source code of their smart contract to Etherscan~\cite{etherscan} in order to facilitate the detection of their ``vulnerable" smart contract.
Etherscan is an online platform that allows anyone to navigate through the Ethereum blockchain and to publish the source code of any smart contract.
Honeypots can be grouped into different techniques, depending on the trapdoor that they use.
Torres et al.~\cite{ferreira2019art} present a tool called \textsc{HoneyBadger} that applies symbolic analysis to contract bytecode and relies on handcrafted rules to detect honeypots.
Unfortunately, this approach cannot recognize new honeypot techniques unless experts are aware of them and specific detection rules are implemented.
Moreover, \textsc{HoneyBadger} only relies on the bytecode of a smart contract and does not consider transaction data. 
However, transaction data reflects user behavior which might be useful to detect honeypot techniques.

In this study, we propose to train a model to classify Ethereum smart contracts into honeypots and non-honeypots. 
Our model takes features from the source code, transaction data and flow of funds. 
We use the results from the \textsc{HoneyBadger} study as labels in order to train our model and validate our results.
To the best of our knowledge, no previous study has analysed the behavior of honeypot contracts
in the Ethereum blockchain from a data science perspective.
In summary, we make three contributions.
First, we present a step by step methodology to obtain, process and explore data related to Ethereum smart contracts and their transactions. 
Second, we propose several honeypot classification models that can generalize well to unseen honeypot instances and techniques.
Third, we describe how we used our machine learning models to discover cases of honeypot contracts that were not detected by \textsc{HoneyBadger}.
We also found honeypots from two completely new techniques that we define in this paper.

\section{Background}

\subsection{Data Science and Machine Learning}

Data science is a discipline that draws techniques from different fields like mathematics, statistics, computer science, and information science, with the goal of discovering insights from real world data.
The term is closely related to other fields such as machine learning, data mining and business intelligence.
Machine Learning is the sub-field of Artificial Intelligence that provides computer programs with the ability to improve from experience without being explicitly programmed~\cite{mitchell1999machine}.
Supervised machine learning algorithms can identify patterns that relate \textit{features} (measurable characteristics of the data) to \textit{labels} (a particular property of the data).
The \textit{learning} or \textit{training} process takes place when the algorithms search for patterns based on examples for which the labels are known.
The result is a \textit{model}, an approximation of the underlying relationship between features and labels.
During the \textit{testing} step, the model assigns labels to samples that were not part of the training phase, and these assignments or \textit{predictions} are compared to known labels.
The purpose is to evaluate how well the model generalizes to unseen cases.
If each label represents one option from a set of possible \textit{classes}, a supervised machine learning problem is known as a \textit{classification}.

XGBoost \cite{Chen:2016:XST:2939672.2939785} is a very popular tool among the data science community, but it is in fact a fast implementation of an older algorithm known as gradient boosted machines~\cite{friedman2001greedy}.
A boosting algorithm is an ensemble of sequential models where each model corrects the errors of the previous one.
In particular, gradient boosting refers to the technique of minimizing the prediction residuals using gradient descent.
The default and most common ensemble members for XGBoost are Decision Trees~\cite{bishop2006pattern}.
Like many other machine learning models, XGBoost outputs for each sample the probability of belonging to each class.
For a binary classification, the predicted class is usually defined by splitting the predicted probability in two using 0.5 as a threshold.

\subsection{Honeypots}


Torres et al.~\cite{ferreira2019art} identified eight different honeypot techniques.
In general, honeypots use a bait in the form of funds to lure users into traps.
After looking at the source code\footnote{The source code is often made publicly available to lure victims.}, users are tricked into believing that there is a vulnerability and that they will receive more funds than they require to trigger the vulnerability.
However, it is not clear to the users that the honeypot contains a hidden trapdoor that causes all or most of the funds to remain in the honeypot.
In the following, we provide a summary of the hidden trapdoors that honeypots use to deceive naive users.


\textit{Balance Disorder (BD).}
This honeypot technique uses the fact that some users are misinformed regarding the exact moment at which a contract's balance is updated.
A honeypot using this technique promises to return its balance and the value sent to it, if the latter is larger than the balance.
However, the sent value is added to the balance before executing the honeypot's code.
As a result, the balance will never be smaller than the sent value.


\textit{Inheritance Disorder (ID).}
This honeypot technique involves two contracts and an inheritance relationship between them.
Both the parent and the child contract each define a variable with the same name that guards the access to the withdrawal of funds.
Although novice users may believe that both variables are the same, they are treated differently by the smart contract.
As a result, changing the content of one variable does not change the content of the other variable.

\textit{Skip Empty String Literal (SESL).}
A bug previously contained in the the Solidity compiler skips hard coded empty string literals in the parameters of function calls.
Thus, the parameters following the empty string literal are shifted upwards.
This can be used to redirect the transfer of funds to the attacker instead of the users.

\textit{Type Deduction Overflow (TDO).}
Although variables are usually declared with specific types, type deduction is also supported\footnote{Until compiler version 0.5.0 \url{https://solidity.readthedocs.io/en/v0.5.0/}.}.
A type is then inferred, without the guarantee that it is sufficiently large.
This can be used to cause integer overflows that prematurely terminate execution and make the honeypot deviate from the behavior expected by the user.

\textit{Uninitialised Struct (US).}
Structs are a convenient way in Solidity to group information.
However, they need to be initialised appropriately, as values written to an uninitialised struct are written by default to the beginning of the storage.
This overwrites existing values and modifies the contract's state.
Thus, variables required to perform the transfer of funds may be overwritten and thereby prevent the transfer.

\textit{Hidden State Update (HSU).}
Etherscan does not display internal transactions that do not transfer any value.
This can be used to hide state updates from users and therefore make them believe that the contract is in a totally different state.

\textit{Hidden Transfer (HT).}
Platforms such as Etherscan allow users to view the source code of published smart contracts.
However, due to the limited size of the source code window, publishers can insert white spaces to hide the displaying of certain source code lines by pushing them outside the window.
This allows publishers to inject and hide code that transfers funds to them instead of users.


\textit{Straw Man Contract (SMC).}
When uploading a contract's source code to Etherscan, only the contract itself is verified and not the code of contracts that the contract may call.
For example, a source code file may contain two contracts, where the first contract makes calls to the second contract.
The second one however, is actually not used after deployment.
Instead, it is a straw man standing in place of another contract that selectively reverts transactions that transfer funds to users.

\section{Data acquisition and feature extraction}

\subsection{Data}


We limited this study to the 2,019,434 smart contracts created in the first 6.5M blocks of the Ethereum blockchain.
Using the Etherscan API~\cite{etherscanapi}, we acquired compilation and source code information for 158,863 contracts, and downloaded around 141M normal and 4.6M internal transactions.

\subsection{Labels}

We compile a list of contracts labeled as honeypots from \textsc{HoneyBadger}'s public code repository~\cite{githubhoneybadger}.
The repository provides a list of honeypots for each technique where source code is available on Etherscan.
The authors developed a tool that generated the labels, and for each case they indicated if the labeling was correct after a manual inspection of the contract source code.
For groups of honeypot contracts that share the exact same byte code, the authors only included one representative address in their list.
In order to obtain a label for every contract, we calculate the \textit{sha256} hash for the byte code of each contract, and group them based on the hashes.
We then transfer the label of each representative contract to every contract in its group.
We only label as honeypot the contracts that were manually confirmed by the authors, and ignore the false positives generated by their tool.

\subsection{Features}

We create a set of features in order to obtain a better understanding of the data, to verify that the behavior of transactions reflects the used honeypot technique, and to automate the detection of honeypots using machine learning.

We divide the features into three categories, depending on where they originate from: source code information, transaction properties, and fund flows.
In the following sections we describe the extraction process for each of the features in each category and the reasons for their selection.

\subsubsection{Source Code Features}

\begin{itemize}
    \item \textit{hasByteCode:bool} $\rightarrow$ indicates if the contract was created with bytecode. The absence of bytecode means that the contract creation was erroneous and that the contract has no behavior.
    
    \item \textit{hasSourceCode:bool} $\rightarrow$ we hypothesize that the number of source code lines of a honeypot is a decent approximation of the contract's readability or interpretability.
    
    \item \textit{numSourceCodeLines:int} $\rightarrow$ counting the number of lines in the source code does not entirely reflect readability and interpretability characteristics, but it is a decent approximation that does not require static code analysis or symbolic execution.
    
    \item \textit{compilerRuns:int} $\rightarrow$ we are interested in determining if a high amount of optimizations during the compilation could lead to an unexpected behavior that can be exploited by honeypots.
    
    \item \textit{library:int} $\rightarrow$ we collected all the possible libraries and created a categorical variable of 749 unique values.
    We are interested in knowing if a library in particular is correlated with the occurrence of honeypots.
\end{itemize}

\noindent And finally, we split the \textit{compilerVersion:str} into three parts: major, minor and patch. We are interested in finding if a major, minor or patch version in particular affected the occurrence of honeypots, caused by some change in the solidity language compilation that can be exploited. For each part, we collected all the possible values and calculated two categorical variables:

\begin{itemize}
    \item \textit{compilerMinorVersion:int} $\rightarrow$ 4 unique values.
    \item \textit{compilerPatchVersion:int} $\rightarrow$ 330 unique values.
\end{itemize}

\noindent The compiler major version is 0 for all the examined contracts, so we discard the feature for not providing any information.

\subsubsection{Transaction Features}
\label{subsec:transaction-features}

We begin with the features we generate for both normal and internal transactions.

\begin{itemize}
    \item \textit{[internal\text{\textbar}normal]TransactionCount:int} $\rightarrow$ the number of transactions of the corresponding type associated with the contract. We expect to filter out contracts with an abundance of movements, which are probably not honeypots.
    
    \item \textit{[internal\text{\textbar}normal]TransactionOtherSenderRatio:float} $\rightarrow$ we calculate this by dividing the number of unique senders by the number of total senders, omitting the contract creator in both.
    The feature should be 1 if every transaction is sent by a different sender, close to zero if few senders send many transactions and we define it as zero if the creator is the only sender.
    We would like to know if the participation of multiple senders is a characteristic of honeypots or not.
    
\end{itemize}

\noindent Next, for each contract and for each property of transactions involving fund movements, we compute the average and the standard deviation.
This yields a total of six features per transaction type, normal and internal:

\begin{itemize}
    \item \textit{[internal\text{\textbar}normal]TransactionValue[Mean\text{\textbar}Std]:float}
    \item \textit{[internal\text{\textbar}normal]TransactionGas[Mean\text{\textbar}Std]:float}
    \item \textit{[internal\text{\textbar}normal]TransactionGasUsed[Mean\text{\textbar}Std]:float}
\end{itemize}

\noindent The value is only collected if there are no errors in the transaction. If there is an error, the transaction including any transferred values is rolled back, but the gas is still spent.
We expect the feature means to be positive but small for honeypots, more varied for popular non-honeypot contracts, and zero for contracts with testing purposes.
Furthermore, for the feature standard deviations we have some weak intuitions we would like to test: high standard deviations for gas and gas used may be correlated with contracts with many different execution paths, becoming a proxy indicator for the complexity of the contract code.
The standard deviation for the value may point out that the contract deals with different amounts of funds.

Moreover, we extract features related to time and block number for normal transactions.
Internal transactions do not carry their own timestamps or block numbers, and we cannot assume the results from Etherscan to be chronologically sorted.

\begin{itemize}
    \item \textit{normalTransactionBlockSpan:int} $\rightarrow$ derived from the block numbers of the first and last transaction.
    
    \item \textit{normalTransactionTimeSpan:int} $\rightarrow$ computed from the timestamps of the first and last transaction.
    
\end{itemize}

\noindent Our intuition is that honeypots have short lives compared to other contracts, which was also pointed out in~\cite{ferreira2019art}.

We also compute the difference of block and time between transactions. Both the mean and standard deviation indicate how often the contract receives transactions (only for contracts with more than one transaction):

\begin{itemize}
    \item \textit{normalTransactionBlockDelta[Mean\text{\textbar}Std]:float}
    \item \textit{normalTransactionTimeDelta[Mean\text{\textbar}Std]:float}
\end{itemize}

Lastly, features extracted only for internal transactions:

\begin{itemize}
    \item \textit{internalTransactionCreationCount:int} $\rightarrow$ indicates how many contracts this contract created.
    Our hypothesis is that honeypots are less likely to create other contracts.
    
    \item \textit{internalTransactionToOtherRatio:float} $\rightarrow$ this is analogous to the feature \textit{transactionOtherSenderRatio}, however, it is calculated for the receivers instead of for the senders.
\end{itemize}

\subsubsection{Fund Flow Features}
\label{subsubsec:fund-flow-features}

Finally, we define fund flow features, based on the definition of honeypots.
Accordingly, we expect the following fund flow events to occur frequently with honeypots.

\begin{itemize}
    \item[1.] A valid contract creation.
    
    \item[2.] An initial deposit, that is, the creator sending ether to the contract.
    This could be interpreted as ``setting the trap", because without funds, no attacker would be interested in exploiting flaws in the contract.
    
    \item[3.] A victim falling for the trap, which means, a transaction sent by an account different than the creator, for which the funds are kept in the contract.
    
    \item[4.] A withdraw of the profit to the creator.
\end{itemize}

\noindent Most importantly, a honeypot should rarely, if ever, present a withdrawal from an account other than its creator.

We define different events triggered by normal transactions (and potentially followed by internal transactions), where each possible event corresponds to exactly one case.
We describe these cases with the combination of eight variables in Table \ref{table:fund-flow-variables}.

\begin{table}[ht]
\centering
\caption{Fund Flow Variables.}
\label{table:fund-flow-variables}
\begin{tabular}{|l|l|}
\hline
\textbf{Name} & \textbf{Possible Values} \\
\hline
\textit{sender} & creator, other \\
\textit{creation} & yes, no \\
\textit{error} & yes, no \\
\textit{balanceCreator} & up, unchanged, down \\
\textit{balanceContract} & up, unchanged, down \\
\textit{balanceSender} & up, unchanged, down \\
\textit{balanceOtherPositive} & yes, no \\
\textit{balanceOtherNegative} & yes, no\\
\hline
\end{tabular}
\end{table}

For each normal transaction we collect the resulting internal transactions.
The \textit{sender} variable is determined by comparing the transaction source to the contract creator.
The \textit{creation} variable is set to \textit{yes} only for the creation transaction.
However, the \textit{error} variable value in this case is an aggregation: it is set as \textit{True} if the normal transaction or any of the triggered internal transactions contain an error.
For the balance changes, we use the following algorithm: we initialize every account with zero funds moved, and for each transaction (both normal and internal), we subtract the value from the \textit{transaction.from} account and add it to the \textit{transaction.to} account (or to the \textit{transaction.contractAddress} account if a contract is created).
After iterating over the transactions, we consider the funds moved into the \textit{creator}, \textit{contract} and \textit{sender} accounts. We mark them as \textit{up}, \textit{unchanged} or \textit{down} if the values are positive, zero or negative respectively.
If the balance of any account other than the three mentioned above is changed, we have two special flags that indicate if at least one other account balance went up or down respectively.
We prefer to keep a small and fixed number of variables and computing all the possible combinations yields $2^5 \times 3^3 = 864$ cases, many of them invalid.
For example, \textit{creation=yes} only makes if when \textit{sender=creator}, and we only care about \textit{balanceSender} if \textit{sender=other}.
Also, at least one balance goes up if and only if at least one balance goes down.
After discarding invalid combinations, we retain $244$ valid cases.
Finally, for every contract, we extract the frequency of each case, by counting how many times a case appears with respect to a contract divided by the total amount of cases related to it.
\section{Exploratory Analysis}
\label{sec:analysis}


\subsection{Most Relevant Numerical Features}

We take all the numerical features and study their distributions for both classes, honeypots and non-honeypots to see if there is a clear difference.
In this study, we include only the results for the features we consider the most relevant.
This is based on their difference with respect to the two classes.


First, we observe that honeypots seem to have a reasonable amount of lines of code in general, with a minimum that could include a fair amount of logic, and a maximum that is still readable.
On the other side, non-honeypots present extreme values, from contracts with only one line of code to a contract with more than 11K lines.
Second, we see that three quarters of the contracts have a small number of normal transactions.
However, the maximum amount is much larger for non-honeypots than for honeypots.
Lastly, it seems that all honeypots always move some amount of ether, but most non-honeypot contracts (over 75\%) do not exchange value at all.
Nonetheless, non-honeypots have very extreme maximum value movements, while honeypots move on average no more than 2 ether.

Handcrafted rules could be used to cut off the extremes of all these features.
However, one of the problems of those rules, is that they are overfitted to the dataset used to craft them.
Even if the boundaries are relaxed, if a new honeypot crosses one of those boundaries (both accidentally or on purpose), it would be automatically classified as non-honeypot disregarding the rest of its properties.
In any case, the rules we could craft can capture all the honeypots but do not exclude a sufficiently large portion of non-honeypots.
We could start combining several rules until we find the best honeypot filter, but the search for the best criteria is exactly what machine learning is used for.

\subsection{Categorical Variables}

All the categorical features we extract belong to the compilation and source code information category.

The compiler minor version has only four possible values.
Most of the contracts belong to the minor version 0.4, and in particular, all but one honeypot among our samples belong to that version too.
These statistics agree with a popularity peak in Ethereum described by the authors of \textsc{HoneyBadger} \cite{ferreira2019art}.

The compiler patch version has 330 possible values.
One compiler version (0.4.19+commit.c4cbbb05) has been used much more than others to compile honeypots, but there is no specific honeypot technique that is significantly more frequent than others for it.
We repeat the procedure of counting honeypot techniques for the rest of the compiler versions obtaining the same results.
This suggests that honeypot activity is not linked to any compiler version. Rather it seems that the high frequency for some versions is related to the time when they were active, which could be at the same time linked with Ethereum's popularity.

Finally, none of the honeypots have a library.
Hence, we can presume that libraries are not linked at all with honeypot development, or that no library contains vulnerabilities that can be exploited for honeypot related purposes.

\subsection{Fund Flow Cases}




We study the events related to fund movements, in order to verify if honeypots behave as we expect.
For this, we query our data by partially defining some of the fund flow variables and adding the frequencies of all matching cases.


We assumed that honeypots should have a deposit from its creator to work, since funds are needed to attract attackers.
However, we find that 29 honeypots (around 10\% of them) do not match this criteria.
Inspecting individual cases, we find that these contracts are actually honeypots, but that they rely on initial deposits from victims to attract even more victims.
Another explanation could be that the creators of honeypots sometimes use a different account to set their trap.
We do not find clear evidence of such practice among our data, but future work could address this topic, for example, using graph techniques to find colluding accounts.

We find 192 honeypots that do not have any deposits from non-creator accounts, which means that only 103 (roughly 35\% of them) were able to lure victims into sending funds. This metric is close the statistics presented in \cite{ferreira2019art}.
Moreover, only 108 honeypot contracts ($\sim$37\% of them) present withdraws from the creator.
It makes sense to expect the creator to withdraw funds only if the trap was effective.
Yet, the amount of honeypots with creator withdraws is higher than the amount of effective ones, which indicates that the creator sometimes withdraws the bait when the trap seems not to be working.

Honeypots should never present withdraws from non-creator accounts, but 50 cases ($\sim$17\%) matching this criteria.
We find some interesting cases among this group, where the honeypot creator failed to set the trap properly.
For example, one contract\footnote{address=0xbC272B58E7cD0a6002c95aFD1F208898D756C580} that belongs to the Hidden State Update category.
For this kind of honeypot, a transaction is needed to change the value of one or more variables in the contract.
This update mechanism is usually hard to track, because Etherscan does not publish internal transactions that carry no value, and victims fall into a trap when they do not know the state of the contract.
However, for our contract example, the hidden state change transaction is missing, and the users that were supposed to be victims made a profit.

Among the top five most frequent balance movement cases, we find creator deposits and withdrawals, deposits from non-creator accounts, and the contract creation (without any fund movement).
The unexpected top case for honeypots contracts are transactions from non-creator accounts that do not generate any kind of fund movement.
Furthermore, deposits from non-creator accounts are also very common for non-honeypot contracts.
The details are enumerated in Table \ref{table:fund-flow-frequencies}.

\begin{table*}[ht]
\centering
\caption{The five most frequent fund flow cases by binary label, honeypot or non-honeypot. The count represents the number of normal transactions that match the fund flow case. The IDs are arbitrary values used to identify the features later on. For readability purposes the boolean variables with False values and the balance variables without changes are not included.}
\label{table:fund-flow-frequencies}
\begin{tabular}{|r|r|l|}
\hline
\multicolumn{3}{|c|}{Honeypots} \\
\hline
Count & ID & Description \\
\hline
341 & 205 & \textit{sender=other} \\
304 & 83 & \textit{sender=creator, balanceCreator=negative, balanceContract=positive} \\
292 & 33 & \textit{sender=creator, creation=True} \\
168 & 201 & \textit{sender=other, balanceContract=positive, balanceSender=negative} \\
136 & 73 & \textit{sender=creator, balanceCreator=positive, balanceContract=negative} \\
\hline
\multicolumn{3}{|c|}{Non-Honeypots} \\
\hline
Count & ID & Description \\
\hline
94475857 & 205 & \textit{sender=other} \\
9667149 & 207 & \textit{sender=other, balanceSender=negative, balanceOtherPositive=True} \\
9130582 & 201 & \textit{sender=other, balanceContract=positive, balanceSender=negative} \\
8569634 & 77 & \textit{sender=creator} \\
5647532 & 127 & \textit{sender=other, error=True} \\
\hline
\end{tabular}
\end{table*}
\section{Experiments}
\label{sec:experiments}

\subsection{Additional Pre-Processing}

Before starting any of the machine learning experiments, we need to take further data cleaning steps.
Missing values in our dataset are mostly a product of our own feature extraction.
For any kind of transaction, the value aggregated features cannot be calculated when all the transactions from the contract contain errors, because no funds were actually moved.
For normal transactions in particular, the difference of time or block number between consecutive normal transactions cannot be measured when the contract only has one normal transaction.
The rest of the normal transaction features are always defined, because there is always at least one normal transaction (the creation, even if it contains errors).
Nevertheless, for all these situations we found it reasonable to fill the missing values with zeros.
However, a contract may not have any internal transactions and, consequently, all the related aggregated features are not defined.
Even if those features could be filled with zeros, only 18\% of the contracts have internal transactions.
Hence, for the following experiments we discard the aggregated features related to internal transactions, but we add a new boolean feature that indicates the presence of internal transactions.

Furthermore, not every contract is useful for our classification task.
Contracts without bytecode will never be honeypots because they cannot be executed, hence, we discard them.
We also discard contracts without source code because:

\begin{itemize}
    \item It is harder to lure victims looking for vulnerabilities in contracts if the source code is not publicly available.
    
    \item We only have labels for contracts with source code.
    
    \item We cannot compute features from the compilation and source code category for contracts without source code.
\end{itemize}

\noindent After the filtering step, the amount of samples in our dataset is reduced to 158,568 non-honeypots and 295 honeypots.

In addition we find that not all fund flow cases are useful.
Therefore, we discard all those that do not have any frequency across our reduced dataset.
Thus, we end up with 74 out of 244 cases that are useful for describing the transaction behavior.
We also check the variance of each individual feature, but none of them present a variance very close to zero, thus we do not discard any further dataset columns.

Our last step is taking any ``unbounded" numerical feature (the ones that are neither frequencies nor ratios), and apply min-max scaling to shrink them into the range $[0, 1]$.
We use this technique to have every feature in the same range, which allows us to better understand feature importance later on.

\subsection{Honeypot Detection}

In this section, we train a machine learning model called XGBoost \cite{Chen:2016:XST:2939672.2939785} to classify contracts into honeypots (the positive class) or non-honeypots (the negative class).

We use k-fold cross-validation to ensure that the models generalize to unseen data.
The data is split into $k=10$ separate subsets called folds, where one is selected for testing, while the algorithm is trained on all the remaining folds.
The procedure is repeated $k$ times so that each fold is used once for testing.
In particular, we use stratified k-fold cross-validation, in which the folds are selected to preserve class proportions.

Furthermore, we deal with an imbalanced classification problem: our dataset contains many more non-honeypot contracts than honeypot contracts.
This issue has a negative impact in the classification capabilities of machine learning algorithms.
To address this issue, we configure XGBoost to train with a scaling weight for the positive class~\cite{XGboostImbalaced}.

We use the AUROC (Area Under the Receiver Operating Characteristics) to quantify the power of our model on each fold.
The ROC curve describes the TPR (True Positive Rate) against the FPR (False Positive Rate) for different thresholds over the classification probability.
The area under that curve represents how well the model can distinguish the two classes.

\begin{table}[ht]
\centering
\caption{Honeypot classification experiments for different categories of features. We show the AUROC mean and standard deviation for train and test collected across all folds.}
\label{table:experiments}
\begin{tabular}{|l|c|c|}
\hline
Features & Train & Test \\
\hline
All & $0.985 \pm 0.002$ & $0.968 \pm 0.015$ \\
Only Transactions & $0.966 \pm 0.004$ & $0.954 \pm 0.030$ \\
Only Source Code & $0.953 \pm 0.002$ & $0.942 \pm 0.025$ \\
Only Fund Flow & $0.952 \pm 0.002$ & $0.938 \pm 0.023$ \\
\hline
\end{tabular}
\end{table}

In Table~\ref{table:experiments} we compare train and test AUROC for four sets of features: the features based only on source code information, based only on aggregated transaction data, based only on fund flow case frequencies, or using all the features together.
In general, for every experiment the difference between train and test is not large, indicating that all the models can generalize well to unseen samples.
In particular, each individual set of features presents good results, with a slight advantage when using all the feature sets together.
In order to understand better the difference between the four alternatives, we list the three most important features per feature set in Table~\ref{table:feature-importances}.
First, the most important feature from the source code group is the number of source code lines, and the following most important are related to the compiler version.
The problem with relying on the compiler version, is that even if some values are correlated with honeypots in the present dataset, those relations might become irrelevant for new compiler versions.
The compiler versions we cover in this study can still be used in the future, but might become less popular over time.
Second, the frequency of deposits to a contract from its creator (\textit{fundFlowCase83}) is a very distinctive fund flow feature, and the mean value of normal transactions is the most relevant of the transaction aggregated features.
The problem of both categories is that they rely only on transaction information.
Hence, when a contract did not execute any transactions, there is no information to make a classification.
Lastly, the advantage of using all the feature groups together, is that while the most important features from each group are at the top of the joint importance ranking, they also cover the weaknesses of each set.



\begin{table}[ht]
\centering
\caption{The three most important features per feature category.}
\label{table:feature-importances}
\begin{tabular}{|l|l|r|}
\hline
Category & Feature & Importance \\
\hline
All & fundFlowCase83 & 0.657 \\
All & normalTransactionValueMean & 0.107 \\
All & numSourceCodeLines & 0.071 \\
\hline
Only Transactions & normalTransactionValueMean & 0.576 \\
Only Transactions & normalTransactionValueStd & 0.117 \\
Only Transactions & normalTransactionGasUsedStd & 0.077 \\
\hline
Only Source Code & numSourceCodeLines & 0.424 \\
Only Source Code & compilerPatchVersion136 & 0.129 \\
Only Source Code & compilerPatchVersion125 & 0.070 \\
\hline
Only Fund Flow & fundFlowCase83 & 0.799 \\
Only Fund Flow & fundFlowCase201 & 0.036 \\
Only Fund Flow & fundFlowCase79 & 0.032 \\
\hline
\end{tabular}
\end{table}


\subsection{Simulated Detection of Unknown Honeypot Techniques}

The previous experiment measures how well the machine learning model generalizes the classification to contracts that were not seen during training, but those contracts share honeypot techniques with other contracts that were present in the training set.
In the following experiment, we intend to simulate what would happen if we need to classify a contract that belongs to a honeypot technique that we never encountered before.
In order to do so, we change the cross validation strategy by picking one honeypot technique at a time: we build the testing set only with the samples belonging to it, and the training set with the remaining samples.
This results in the eight different experiments that are shown in Table~\ref{table:experiments-one-vs-all}.
Because the test set is composed solely of positive cases, we measure the test $Recall = TP / (TP + FN)$.

\begin{table}[ht]
\centering
\caption{Simulating the Detection for Unknown Honeypot Techniques. The test step includes only honeypots from one honeypot type at a time, and the train involves the rest of the contracts.}
\label{table:experiments-one-vs-all}
\begin{tabular}{|l|c|c|c|}
\hline
Removed Honeypot Technique & FN & TP & Recall \\
\hline
Type Deduction Overflow & 0 & 4 & 1.000 \\
Uninitialised Struct & 2 & 37 & 0.949 \\
Hidden Transfer & 1 & 12 & 0.923 \\
Hidden State Update & 12 & 123 & 0.911 \\
Inheritance Disorder & 4 & 39 & 0.907 \\
Straw Man Contract & 3 & 28 & 0.903 \\
Skip Empty String Literal & 1 & 9 & 0.900 \\
Balance Disorder & 3 & 17 & 0.850 \\


\hline
\end{tabular}
\end{table}

The high recall for all cases implies that the machine learning approach can detect honeypots that were not seen during training.
This is a clear advantage over \textsc{HoneyBadger}.
Because it relies on expert knowledge of the different honeypot techniques to craft detection rules, it is unable to detect new, yet unknown techniques.

\subsection{Detection of New Honeypots and Honeypot Techniques}
\label{sec:newhpots}

The labels we obtained from \textsc{HoneyBadger} correspond to contracts that the tool marked as honeypots and that were afterwards manually validated by its authors.
Nevertheless, it does not mean that the rest of the contracts are necessarily non-honeypots.
A considerable amount of human work is required to inspect millions of instances of Solidity source code.
In this section we describe a mechanism to review the unlabeled contracts in a time efficient manner.
Firstly we train ten different XGBoost models using ten stratified folds, and for each contract we calculate the probability of being a honeypot with each model.
Secondly, we compute the mean and standard deviation of the ten values for each contract.
Lastly, we rank the contracts in descending order using the mean probability of being a honeypot.
A large standard deviation can be used to filter out the contracts for which the models did not agree on the prediction.
The resulting ranking serves as a guide to the order in which the contracts should be inspected.
We decided to manually validate the first 100 samples of the list, and discovered 57 new instances of honeypots.
The amount of honeypot instances by technique are described in Table~\ref{table:new-honeypots}.
From those instances, we found 3 extra copies by matching hashes of contract byte code.
Among the techniques of the honeypots we found, two are not part of the \textsc{HoneyBadger} study.

\begin{table}[ht]
\centering
\caption{Number of new honeypots discovered by technique.}
\label{table:new-honeypots}
\begin{tabular}{|c|c|c|c|c|c|c|c|c|}
\hline
BD & ID & TDO & US & HSU & HT & SMC & UC & MKET \\
\hline
1 & 13 & 1 & 6 & 16 & 8 & 5 & 6 & 1 \\
\hline
\end{tabular}
\end{table}

The first of the newly found honeypots uses a \textit{call} to transfer funds to a potential attacker, provided they sent funds first.
However, the \textit{call} is not complete: it is prepared, but never executed, as the parenthesis required for the execution are missing.
\textsc{Honeybadger} cannot detect this honeypot, as most compiler versions remove the unnecessary bytecode that sets up the \textit{call}.
Furthermore, since the \textit{call} is never executed, there is also no \textit{CALL} opcode within the bytecode.
With the approach proposed in this paper, however, this honeypot technique can be detected, as more information, such as transactions sent to contracts, is analyzed.
We propose to name this honeypot technique \textit{Unexecuted Call (UC)}.

The second honeypot\footnote{For example: 0xf5615138a7f2605e382375fa33ab368661e017ff.} that has been found with our approach is detailed in the following.
It implements transfers to the owner only.
As such it seems that the honeypot serves as a container for the owner's funds.
The owner however can seemingly be overwritten in a function that serves to initialize the contract and can be called by anyone.
The honeypot suggests that anyone can become owner and then transfer all contained funds to themselves.
However, the contract owner can never truly be changed from the value set during the deployment.
This is due to the contract making use of a mapping to store the owner variable.
This mapping uses a string, "Stephen", as the key to access the owner.
When writing the contract owner during deployment and whenever reading the owner variable, the correct key is used.
However, in the initialization function, a misleading key is employed.
Instead of using the same key when overwriting the owner, the used key is slightly, but not noticeably different.
In fact, a cyrillic letter "e"\footnote{Unicode code 1077: \url{https://www.codetable.net/decimal/1077}.} is introduced as the second "e" in "Stephen".
It is so similar to the basic "e"\footnote{Unicode code 101: \url{https://www.codetable.net/decimal/101}.} that a user cannot easily recognize it.
Nevertheless, the code makes the distinction.
Therefore, the initialization function, which requires to be sent funds to in order to be executed properly is unable to update the owner.
Instead, it stores the new value under an altogether different key.
Moreover, only the real owner can retrieve the funds.
Due to the employed technique, we propose to name this honeypot technique \textit{Map Key Encoding Trick (MKET)}.

\section{Related Work}











In \cite{ferreira2019art}, Torres et al. proposed the first taxonomy of honeypots.
They identified eight different honeypot techniques and implemented a tool, \textsc{HoneyBadger}, to detect them.
\textsc{Honeybadger} uses static analysis to analyse Ethereum smart contracts.
Before that, in the domain of smart contracts, static analysis has been used to detect vulnerabilities \cite{Luu2016, torres2018osiris}, and to verify the compliance and violation of patterns \cite{tsankov2018securify}.
While other such tools exist, they focus on the security of smart contracts and not on honeypots or scams.
Scams, specifically Ponzi schemes, on Ethereum are investigated in \cite{bartoletti2017dissecting}, however smart contracts with source code are analyzed manually.
A tool for detecting Ponzi schemes using machine learning is discussed in \cite{chen2018detecting}.
Machine learning has also been used in \cite{tann2018towards} to rapidly adapt to new attack patterns in order to detect them.
Other uses of machine learning in this domain put more emphasis on transactions, however also for security reasons \cite{nikolic2018finding} and \cite{krupp2018teether}, to detect anomalies \cite{camino2017finding} or to predict the bitcoin price \cite{mcnally2018predicting}.

In this paper, we use a machine learning approach to detect a specific type of scam, honeypots.
Similar to other works, we use source code information and transaction data.
However, we also put emphasis on the flow of funds.
Our machine learning approach, as opposed to \textsc{HoneyBadger}, also allows us to discover new honeypots, as discussed in Section \ref{sec:newhpots}.





\section{Future Work}

There are several ways in which this paper can be extended.
We could explore the possibility of extracting features from the contract names based on the work presented in \cite{norvill2017automated}, by crossing the occurrence of words in the contract names with a list of relevant terms for the Ethereum community.
Another alternative could be to calculate a global state of account balances after the execution of all the transactions of a contract, instead of only checking fund flows per transaction.

We conceived the fund flow cases with the initial goal of analysing the temporal dependencies between each other.
In other words, we wanted to study the behavior of contracts using sequences of discrete symbols, which is equivalent to work with character strings or list of words.
However, as we found that most of the honeypots present very short sequences, the approach was not successful in finding useful patterns.
In the end, we only used in this paper uni-gram frequencies, and discarded other alternatives like Markov Chains, n-gram frequencies, or Recurrent Neural Networks.
Nevertheless, we plan to employ fund flow case sequences in future studies, to classify contracts (or even accounts) with a larger transaction history, into categories that do not involve honeypots.
\section{Conclusion}

In this paper, we presented a step by step methodology to obtain, process and analyze Ethereum contracts for the task of honeypot detection.
We showed how assumptions and hypotheses about honeypot behavior can be contrasted with real data, and derived features for classification models\footnote{Source code available at \url{https://github.com/rcamino/honeypot-detection}.}.
The machine learning models proved to generalize well, even when all the contracts belonging to one honeypot technique were removed from training.
Most importantly, we showed that our technique detected honeypots from two new techniques, which would be impossible to achieve using byte code analysis without manually crafting new detection rules.
Furthermore, we proved that \textsc{HoneyBadger}'s rules covering known honeypot techniques were not capturing all the existing honeypot instances.





\bibliographystyle{IEEEtran}
\bibliography{bibliography}

\begin{thebibliography}{10}
\providecommand{\url}[1]{#1}
\csname url@samestyle\endcsname
\providecommand{\newblock}{\relax}
\providecommand{\bibinfo}[2]{#2}
\providecommand{\BIBentrySTDinterwordspacing}{\spaceskip=0pt\relax}
\providecommand{\BIBentryALTinterwordstretchfactor}{4}
\providecommand{\BIBentryALTinterwordspacing}{\spaceskip=\fontdimen2\font plus
\BIBentryALTinterwordstretchfactor\fontdimen3\font minus
  \fontdimen4\font\relax}
\providecommand{\BIBforeignlanguage}[2]{{%
\expandafter\ifx\csname l@#1\endcsname\relax
\typeout{** WARNING: IEEEtran.bst: No hyphenation pattern has been}%
\typeout{** loaded for the language `#1'. Using the pattern for}%
\typeout{** the default language instead.}%
\else
\language=\csname l@#1\endcsname
\fi
#2}}
\providecommand{\BIBdecl}{\relax}
\BIBdecl

\bibitem{nakamoto2008bitcoin}
S.~Nakamoto, ``Bitcoin: A peer-to-peer electronic cash system,''
  \emph{Cryptography Mailing list at https://metzdowd.com}, 03 2009.

\bibitem{wood2014ethereum}
G.~Wood, ``Ethereum: A secure decentralised generalised transaction ledger,''
  \emph{Ethereum Project Yellow Paper}, vol. 151, pp. 1--32, 2014.

\bibitem{solidity}
------, ``Solidity 0.5.11 documentation,'' September 2019,
  https://solidity.readthedocs.io/en/v0.5.11/.

\bibitem{atzei2017}
N.~Atzei, M.~Bartoletti, and T.~Cimoli, ``{A Survey of Attacks on Ethereum
  Smart Contracts (SoK)},'' in \emph{Proceedings of the 6th International
  Conference on Principles of Security and Trust - Volume 10204}.\hskip 1em
  plus 0.5em minus 0.4em\relax Springer-Verlag New York, Inc., 2017, pp.
  164--186.

\bibitem{siegel2016understanding}
D.~Siegel, ``Understanding the dao attack,'' jun 2016,
  https://www.coindesk.com/understanding-dao-hack-journalists/.

\bibitem{petrov2017another}
S.~Petrov, ``Another parity wallet hack explained,'' nov 2017,
  https://medium.com/@Pr0Ger/another-parity-wallet-hack-explained-847ca46a2e1c.

\bibitem{ferreira2019art}
C.~Ferreira~Torres, M.~Steichen \emph{et~al.}, ``The art of the scam:
  Demystifying honeypots in ethereum smart contracts,'' in \emph{USENIX
  Security Symposium, Santa Clara, 14-16 August 2019}, 2019.

\bibitem{etherwars}
D.~Luca and B.~Mueller, ``{The Ether Wars: Exploits, counter-exploits and
  honeypots on Ethereum},'' August 2019,
  https://infocondb.org/con/def-con/def-con-27/the-ether-wars-exploits-counter-exploits-and-honeypots-on-ethereum.

\bibitem{hacking_the_hackers}
A.~Sherbachev, ``Hacking the hackers: Honeypots on ethereum network,'' December
  2018,
  https://hackernoon.com/hacking-the-hackers-honeypots-on-ethereum-network-5baa35a13577.

\bibitem{honeypot_analysis}
J.~Sanjuas, ``An analysis of a couple ethereum honeypot contracts,'' December
  2018,
  https://medium.com/coinmonks/an-analysis-of-a-couple-ethereum-honeypot-contracts-5c07c95b0a8d.

\bibitem{etherscan}
Etherscan, ``Etherscan,'' December 2019, \url{https://etherscan.io/}.

\bibitem{mitchell1999machine}
T.~M. Mitchell, ``Machine learning and data mining,'' \emph{Communications of
  the ACM}, vol.~42, no.~11, 1999.

\bibitem{Chen:2016:XST:2939672.2939785}
\BIBentryALTinterwordspacing
T.~Chen and C.~Guestrin, ``{XGBoost}: A scalable tree boosting system,'' in
  \emph{Proceedings of the 22nd ACM SIGKDD International Conference on
  Knowledge Discovery and Data Mining}, ser. KDD '16.\hskip 1em plus 0.5em
  minus 0.4em\relax New York, NY, USA: ACM, 2016, pp. 785--794. [Online].
  Available: \url{http://doi.acm.org/10.1145/2939672.2939785}
\BIBentrySTDinterwordspacing

\bibitem{friedman2001greedy}
J.~H. Friedman, ``Greedy function approximation: a gradient boosting machine,''
  \emph{Annals of statistics}, pp. 1189--1232, 2001.

\bibitem{bishop2006pattern}
C.~M. Bishop, \emph{Pattern recognition and machine learning}.\hskip 1em plus
  0.5em minus 0.4em\relax springer, 2006.

\bibitem{etherscanapi}
Etherscan, ``Ethereum developer apis,'' December 2019,
  \url{https://etherscan.io/apis}.

\bibitem{githubhoneybadger}
C.~Ferreira~Torres, ``christoftorres/honeybadger,'' December 2019,
  \url{https://github.com/christoftorres/HoneyBadger}.

\bibitem{XGboostImbalaced}
xgboost developers, ``Notes on parameter tuning - handle imbalanced dataset,''
  December 2019,
  \url{https://xgboost.readthedocs.io/en/latest/tutorials/param_tuning.html#handle-imbalanced-dataset}.

\bibitem{Luu2016}
L.~Luu, D.-H. Chu, H.~Olickel, P.~Saxena, and A.~Hobor, ``Making smart
  contracts smarter,'' in \emph{Proceedings of the 2016 ACM SIGSAC Conference
  on Computer and Communications Security}, ser. CCS '16.\hskip 1em plus 0.5em
  minus 0.4em\relax New York, NY, USA: ACM, 2016, pp. 254--269.

\bibitem{torres2018osiris}
C.~F. Torres, J.~Sch\"{u}tte, and R.~State, ``Osiris: Hunting for integer bugs
  in ethereum smart contracts,'' in \emph{Proceedings of the 34th Annual
  Computer Security Applications Conference}, ser. ACSAC '18.\hskip 1em plus
  0.5em minus 0.4em\relax New York, NY, USA: ACM, 2018, pp. 664--676.

\bibitem{tsankov2018securify}
P.~Tsankov, A.~Dan, D.~Drachsler-Cohen, A.~Gervais, F.~Buenzli, and M.~Vechev,
  ``Securify: Practical security analysis of smart contracts,'' in
  \emph{Proceedings of the 2018 ACM SIGSAC Conference on Computer and
  Communications Security}.\hskip 1em plus 0.5em minus 0.4em\relax ACM, 2018,
  pp. 67--82.

\bibitem{bartoletti2017dissecting}
\BIBentryALTinterwordspacing
M.~Bartoletti, S.~Carta, T.~Cimoli, and R.~Saia, ``Dissecting ponzi schemes on
  ethereum: identification, analysis, and impact,'' \emph{CoRR}, vol.
  abs/1703.03779, 2017. [Online]. Available:
  \url{http://arxiv.org/abs/1703.03779}
\BIBentrySTDinterwordspacing

\bibitem{chen2018detecting}
\BIBentryALTinterwordspacing
W.~Chen, Z.~Zheng, J.~Cui, E.~Ngai, P.~Zheng, and Y.~Zhou, ``Detecting ponzi
  schemes on ethereum: Towards healthier blockchain technology,'' in
  \emph{Proceedings of the 2018 World Wide Web Conference}, ser. WWW '18.\hskip
  1em plus 0.5em minus 0.4em\relax Republic and Canton of Geneva, Switzerland:
  International World Wide Web Conferences Steering Committee, 2018, pp.
  1409--1418. [Online]. Available:
  \url{https://doi.org/10.1145/3178876.3186046}
\BIBentrySTDinterwordspacing

\bibitem{tann2018towards}
\BIBentryALTinterwordspacing
W.~J. Tann, X.~J. Han, S.~S. Gupta, and Y.~Ong, ``Towards safer smart
  contracts: {A} sequence learning approach to detecting vulnerabilities,''
  \emph{CoRR}, vol. abs/1811.06632, 2018. [Online]. Available:
  \url{http://arxiv.org/abs/1811.06632}
\BIBentrySTDinterwordspacing

\bibitem{nikolic2018finding}
\BIBentryALTinterwordspacing
I.~Nikoli\'{c}, A.~Kolluri, I.~Sergey, P.~Saxena, and A.~Hobor, ``Finding the
  greedy, prodigal, and suicidal contracts at scale,'' in \emph{Proceedings of
  the 34th Annual Computer Security Applications Conference}, ser. ACSAC
  '18.\hskip 1em plus 0.5em minus 0.4em\relax New York, NY, USA: ACM, 2018, pp.
  653--663. [Online]. Available:
  \url{http://doi.acm.org/10.1145/3274694.3274743}
\BIBentrySTDinterwordspacing

\bibitem{krupp2018teether}
\BIBentryALTinterwordspacing
J.~Krupp and C.~Rossow, ``teether: Gnawing at ethereum to automatically exploit
  smart contracts,'' in \emph{27th {USENIX} Security Symposium ({USENIX}
  Security 18)}.\hskip 1em plus 0.5em minus 0.4em\relax Baltimore, MD: {USENIX}
  Association, Aug. 2018, pp. 1317--1333. [Online]. Available:
  \url{https://www.usenix.org/conference/usenixsecurity18/presentation/krupp}
\BIBentrySTDinterwordspacing

\bibitem{camino2017finding}
R.~D. Camino, R.~State, L.~Montero, and P.~Valtchev, ``Finding suspicious
  activities in financial transactions and distributed ledgers,'' in \emph{2017
  IEEE International Conference on Data Mining Workshops (ICDMW)}.\hskip 1em
  plus 0.5em minus 0.4em\relax IEEE, 2017, pp. 787--796.

\bibitem{mcnally2018predicting}
S.~McNally, J.~Roche, and S.~Caton, ``Predicting the price of bitcoin using
  machine learning,'' in \emph{2018 26th Euromicro International Conference on
  Parallel, Distributed and Network-based Processing (PDP)}.\hskip 1em plus
  0.5em minus 0.4em\relax IEEE, 2018, pp. 339--343.

\bibitem{norvill2017automated}
R.~Norvill, B.~B.~F. Pontiveros, R.~State, I.~Awan, and A.~Cullen, ``Automated
  labeling of unknown contracts in ethereum,'' in \emph{2017 26th International
  Conference on Computer Communication and Networks (ICCCN)}.\hskip 1em plus
  0.5em minus 0.4em\relax IEEE, 2017, pp. 1--6.

\end{thebibliography}

\end{document}